\documentclass[11pt,a4paper]{article}
\usepackage{amssymb, amsmath,framed}
\usepackage[colorlinks]{hyperref}
\usepackage[margin=1in]{geometry}
\newcommand{\rem}[1]{}
\newcommand{\de}{{\rm d}}
\newcommand{\tf}{{\tilde{f}}}

\newcommand{\bq}{{\mathbf{x}}}
\newcommand{\bc}{{\mathbf{c}}}

\newcommand{\bK}{{\mathbf{K}}}
\newcommand{\bE}{{\mathbf{E}}}
\newcommand{\bA}{{\mathbf{A}}}
\newcommand{\bB}{{\mathbf{B}}}

\newcommand{\ba}{{\mathbf{a}}}

\newcommand{\bu}{{\mathbf{u}}}
\newcommand{\bv}{{\mathbf{v}}}
\newcommand{\bU}{{\boldsymbol{U}}}
\newcommand{\bV}{{\boldsymbol{V}}}
\newcommand{\bW}{{\boldsymbol{W}}}

\newcommand{\bX}{{\mathbf{X}}}
\newcommand{\tX}{{\widetilde{\mathbf{X}}}}

\newtheorem{remark}{Remark}

\begin{document}

\title{A Lagrangian kinetic model for collisionless magnetic reconnection}
\author{Cesare Tronci\\
\it\footnotesize Department of Mathematics, University of Surrey, Guildford GU2 7XH, United Kingdom\\
%\it\footnotesize $^2$Section de Math\'ematiques, \'Ecole Polytechnique
%F\'ed\'erale de Lausanne, Switzerland\vspace{-.2cm}
}
\date{}

\maketitle

\bigskip

\begin{abstract}
A new fully kinetic system is proposed for modeling collisionless magnetic reconnection. The formulation relies on fundamental principles in Lagrangian dynamics, in which the {\color{black}inertia of the electron mean flow} is neglected in the expression of the Lagrangian, rather then enforcing a zero electron mass in the equations of motion. This is done upon splitting the electron velocity into its mean and fluctuating parts, so that the latter naturally produce the corresponding pressure tensor. The model exhibits a new Coriolis force term, which emerges from a change of frame in the electron dynamics. Then, if the electron heat flux is neglected,  the strong electron magnetization limit yields a hybrid model, in which the electron pressure tensor is frozen into the electron mean velocity.
\end{abstract}

\vspace{1cm}

\tableofcontents

%\bigskip

\newpage

\section{Introduction}

A well known necessary condition for magnetic reconnection in plasmas is that the magnetic field lines are not frozen into the fluid flow. For example, while ideal MHD (with frozen-in magnetic field) fails to reproduce reconnection, a great advance is provided by resistive MHD, in which the frozen-in condition is broken by a finite resistivity. However, although reliable results are obtained by resistive MHD in several conditions, modeling collisionless reconnection requires extra effort, due to the special nature of the kinetic features underlying this phenomenon. Kinetic theory features first make their appearance  in the dynamics of electrons, whose pressure tensor is often assumed to dominate over the inertia of their mean flow. In addition, the high energy levels of the ions lead to the necessity of a kinetic treatment also for these particles.

\subsection{The kinetic model\label{kin-model}}
In the light of the above arguments, fully kinetic simulations become necessary in modelling collisionless reconnection and they are based on a set of three equations, which may be conveniently written in terms of the ion distribution on phase-space $f_i(\bq,\bv)$, the \emph{relative} electron distribution $\tf_e(\bq,\bc)$ (where $\bc=\boldsymbol{v}-\bV_e$ and $\bV_e$ is the electron mean velocity), and the magnetic induction field $\bB$. This paper proposes the following model for inertialess electron mean flow:
\begin{align}\label{electron-f}
&
\frac{\partial\tf_e}{\partial t}+\big(\bc+\bV_e\big)\cdot\frac{\partial\tf_e}{\partial \bq}+\left[\frac{q_e}{m_e}\Big(\bE+(\bc+\bV_e)\times\!\bB\Big)-\bc\cdot\nabla\bV_e-\bc\times\nabla\times\bV_e\right]\cdot\frac{\partial\tf_e}{\partial \bc}=0
\\\label{ion-f}
&
\frac{\partial f_i}{\partial t}+\bv\cdot\frac{\partial f_i}{\partial \bq}+\frac{q_i}{m_i}\big(\bE+\bv\times\bB\big)\cdot\frac{\partial f_i}{\partial \bv}=0
\\\label{magneticfield}
&
\frac{\partial\bB}{\partial t}
=-\nabla\times\bE
\end{align}
where 
\begin{equation}
\bE=-{\bV_e}\times\bB
+\frac1{q_e n_e}\nabla\cdot\widetilde{\Bbb{P}}_e
\end{equation}
and
\begin{align}
&\label{kinetic-quantities}
n_e=-\frac{q_i}{q_e} \int\! f_i\,\de^3\bv
\,,\qquad
q_e n_e\bV_e={\color{black}\mu_0^{-1}}\nabla\times\bB-q_i\int\!\bv\, f_i\,\de^3\bv
\,,\qquad
\widetilde{\Bbb{P}}_e=m_e\! \int\!\bc\bc\,\tf\,\de^3\bc
\end{align}
%\noindent\comment{recall to insert back the constants!}\noindent
so that $n_e=\int\!\tf_e\,\de^3\bc$ is the electron density, $\widetilde{\Bbb{P}}_e$ is the electron pressure tensor and $\bE$ is the electric field. Here, the constants $q_{e}$ and $q_i$ denote the electron and ion charge (with sign), respectively, and $m_e$ and $m_i$ denote their corresponding masses. {\color{black} Also, the constant $\mu_0$ is the magnetic constant. While the dynamics of the magnetic field \eqref{magneticfield} and of the} ion probability density \eqref{ion-f} is commonly found in the literature and it is easily derived by neglecting inertial terms in the equation for the electron mean velocity $\bV_e$, the kinetic equation \eqref{electron-f} for the relative electron density exhibits the new Coriolis-type force $\bc\times\boldsymbol\omega_e$, where $\boldsymbol\omega_e=\nabla\times\bV_e$ is the  electron fluid vorticity. This type of non-inertial forcing appears typically in moving frames (recall that $\tf_e$ is the phase-space density in the frame moving with $\bV_e$), as shown in \cite{ThMc} for the case of general magnetized plasmas. However, this term is lost in common models, precisely in the step when one neglects electron inertia, after splitting electron mean and fluctuation velocities. This means that the asymptotic process leading to Hall MHD does not retain important features, such as non-inertial forces, when one aims to account for the fluctuation kinetics. Retaining these important features requires a different method. In particular, equations \eqref{electron-f}--\eqref{kinetic-quantities} are derived in this paper by applying Lagrangian methods: the same methods that yielded  MHD in \cite{HoMaRa1998} and  Hall MHD in \cite{IlLa}.

\subsection{Electron pressure tensor dynamics\label{pressure}}
In a series of papers (see e.g. \cite{WiYiOmKaQu,YiWiGaBi,KuHeWi,KuHeBi,YiWi} and subsequent papers by the same authors), considerable effort has been dedicated to the question of whether a moment truncation is possible, that would give a satisfactory description of  electron kinetics. This process must account for the pressure tensor dynamics, so that the simplest possible truncation would rely on the assumption of a negligible electron heat flux. Under this hypothesis, equation \eqref{electron-f} eventually yields
\begin{multline}\label{pressure-eq}
\frac{\partial\widetilde{\Bbb{P}}_e}{\partial t}+(\bV_e\cdot\nabla)\widetilde{\Bbb{P}}_e+(\nabla\cdot\bV_e)\widetilde{\Bbb{P}}_e+\frac{q_e}{m_e}\left(\bB\times\widetilde{\Bbb{P}}_e-\widetilde{\Bbb{P}}_e\times\bB\right)+\widetilde{\Bbb{P}}_e\cdot\nabla\bV_e+\big(\widetilde{\Bbb{P}}_e\cdot\nabla\bV_e\big)^T
\\{\color{black}
+
\underbrace{\,\widetilde{\Bbb{P}}_e\times\boldsymbol\omega_e
-\boldsymbol\omega_e\times\widetilde{\Bbb{P}}_e\,}_\textit{Coriolis force terms}
}
=0
\,,
\end{multline}
%where $\alpha_e=q_e/m_e$ is the electron charge-to-mass ratio. 
{\color{black}where the superscript $T$ denotes transpose}. Here, the new Coriolis {\color{black}terms persist as they give} non-zero contribution in the pressure tensor dynamics. More importantly, one observes that in the strong electron magnetization limit $\bB\times\widetilde{\Bbb{P}}_e-\widetilde{\Bbb{P}}_e\times\bB\simeq0$, the electron pressure tensor density is \emph{frozen} into the electron mean velocity, that is (upon dropping the index $e$ for convenience of notation)
\begin{equation}\label{advect-correl}
\frac{\de}{\de t}\!\left(\widetilde{\Bbb{P}}_{jk}(\mathbf{q}_t,t)\,\de q^j_t\,\de q^k_t\,\de^3\mathbf{q}_t\right)=0
\,\qquad\text{\ along\ }\qquad
\frac{\de\mathbf{q}_t}{\de t}=\bV_e(\mathbf{q}_t,t)
\,.
\end{equation}
This means that the electron velocity correlations are constant along the electron mean flow. Although, this is a nice picture, we should not forget that it assumes a negligible electron heat flux. The question whether this assumption is justifiable is still open, with results that basically depend on the particular situation that is being considered each time \cite{KuHeBi}.

{\color{black}
\begin{remark}[Pressure tensor density]
It may  be  worth emphasizing that the density $\de^3\mathbf{q}_t$ in  equation \eqref{advect-correl} arises from the fact that the moment definition of the pressure tensor (see third relation in \eqref{kinetic-quantities})
does not involve dividing by the particle density $n_e$, in such a way that the resulting covariant tensor $\widetilde{\Bbb{P}}_e$ retains the spatial density $\de^3\mathbf{x}$. This is a standard fact in the mathematical theory of Vlasov kinetic moments. Then, upon using $\de^3\dot{\mathbf{q}}_t=(\nabla\cdot\bV_e)\,\de^3\mathbf{q}_t$, one can see that \eqref{advect-correl} is equivalent to \eqref{pressure-eq} by a direct verification.
\end{remark}
}

\subsection{Main content of the paper}
\begin{enumerate}
\item 
Section \ref{model-sec} formulates the model by applying the hypothesis of negligible inertia for the electron mean flow. Instead of inserting this hypothesis in the equations of motion, this is done in the variational framework by using standard techniques in the theory of continuum systems with Lagrangian labels \cite{HoMaRa1998}. 

\item Section \ref{elpressure} analyzes the consequences of the model formulated in Section \ref{model-sec} in terms of the electron pressure tensor. It is shown how the strong electron magnetization limit yields the frozen-in condition for the electron pressure tensor, as long as the heat flux contribution may be neglected. Also, it is emphasized how this new frozen-in law for the electron pressure can be used to construct new hybrid models, which discard the information about higher order moments of the electron distribution.

\item Section \ref{morestuff-sec} presents the stationary equations in various cases and shows how the present model recovers the Harris current sheet solutions of the Maxwell-Vlasov system. In the same Section, a conserved total energy is presented explicitly, along with two families of constants of motion that are inherited from the Maxwell-Vlasov system.

\end{enumerate}

\section{Formulation of the model\label{model-sec}}

The Lagrangian approach to continuum systems has a long standing tradition, whose most famous result is probably Arnold's formulation of Euler's fluid equation \cite{Ar66} in terms of geometry and symmetry. Later, this approach was extended to various compressible fluid models in different contexts \cite{HoMaRa1998}. In plasma kinetic theory, the Lagrangian approach appeared in the well known work by Low \cite{Low}, which then was  considered by Dewar in \cite{Dewar}. This approach achieved its most prominent result in Littlejohn's formulation of guiding center motion \cite{Littlejohn}. Later, this approach was pursued by many others \cite{YeMo,Brizard,CeHoHoMa}. When the variational approach arises from an action principle that is written in terms of purely Lagrangian labels, then this approach is often known as Euler-Poincar\'e variational method \cite{HoMaRa1998}. For example, the variational formulations of ideal MHD \cite{HoMaRa1998} and Hall MHD \cite{IlLa} are exactly of this type. {\color{black}Recently, gyrokinetic theory has also been  cast in this framework \cite{SqQiTa}}. The present approach is mainly inspired by the results in \cite{CeHoHoMa,IlLa}. {\color{black}The aim of the following two Sections is to explain how the model \eqref{electron-f}--\eqref{kinetic-quantities} can be derived from a variational principle.}

\subsection{The action principle}
In order to derive an appropriate action principle for modeling collisionless reconnection, let us write the Eulerian action functional as
\begin{multline}\label{actionprinciple}
\delta\!\int_{t_1}^{t_2}\!\left(
\frac{m_i}2\!\int\! f_i\,|{\bu}_i|^2\,\de^3\bq\,\de^3\bv
+q_i\!\int\!f_i{\bu}_i\cdot\bA\,\de^3\bv\,\de^3\bq
-q_i\!\int\!\varphi f_i\,\de^3\bv \,\de^3\bq
\right.
\\
+\frac{m_e}2\!\int \!\tf_e\,|\widetilde{\bu}_e+\epsilon\bV_e|^2\,\de^3\bq\,\de^3\bc
+q_e\!\int\! \tf_e(\widetilde{\bu}_e+\bV_e)\cdot\bA\,\de^3\bc\,\de^3\bq-q_e\!\int\!\varphi \tf_e\,\de^3\bc\,\de^3\bq
\\
\left.-\frac{m_i}2\!\int \!f_i\,|\bu_i-\bv|^2\,\de^3\bq\,\de^3\bv-\frac{m_e}2\!\int \!\tf_e\,|\widetilde{\bu}_e-\bc|^2\,\de^3\bq\,\de^3\bc-\frac1{2{\color{black}\mu_0}}\int|\nabla\times\bA|^2\,\de^3\bq\right)\de t=0
\end{multline}
where the notation is as in equations \eqref{electron-f}--\eqref{kinetic-quantities}. In addition, $(\varphi,\bA)$ denotes the electromagnetic potentials, $\epsilon\in[0,1]$ is a convenient parameter and the quantities
\[
\bu_i=\bu_i(\bq,\bv,t)
\,,\ \qquad\
\widetilde{\bu}_e=\widetilde{\bu}_e(\bq,\bc,t)
\]
denote the position components of the Eulerian phase-space velocity for the two species, so  that $\dot{\mathbf{x}}_i={\bu}_i(\bq,\bv,t)$ for the ions while $\dot{\mathbf{x}}_e=\widetilde{\bu}_e(\bq,\bc,t)+\bV_e(\bq,t)$  for the electrons. {\color{black} Notice that the vector field $\widetilde{\bu}_e$ plays the role of a \emph{fluctuation velocity}, as opposed to the mean velocity $\bV_e$. Under this velocity splitting, the  variational principle \eqref{actionprinciple} is obtained from the one for the Maxwell-Vlasov system \cite{Low,CeHoHoMa,SqQiTa}, upon discarding the electric field energy $\varepsilon_0/2\int\!\left(\nabla\varphi+\partial_t\bA\right)\de^3\mathbf{x}$ ($\varepsilon_0$ being the electric constant).} In more generality,  one defines the six-dimensional phase-space velocities $\bX_i$ and $\widetilde{\bX}_e$ so that, by a slight abuse of notation one may write
\[
(\dot{\mathbf{x}},\dot{\bv})_i=\left(\bu_{i},\mathbf{a}_{i}\right)=:\bX_{i}
\]
for the ions and
\[
(\dot{\mathbf{x}},\dot{\bc})_e=(\widetilde{\bu}_e+\bV_e,\widetilde{\mathbf{a}}_e)
\,,\ \qquad\
\text{with}
\,\ \qquad\
\widetilde{\bX}_e:=(\widetilde{\bu}_e,\widetilde{\mathbf{a}}_e)
\,,
\]
for the electrons. Here, the accelerations $\mathbf{a}_{i}$ and $\widetilde{\mathbf{a}}_e$ do not appear in the action principle \eqref{actionprinciple}, since the total energy cannot depend on the particle accelerations. Along the lines of \cite{CeHoHoMa}, the terms in $|\bu_i-\bv|^2$ and $|\widetilde{\bu}_e-\bc|^2$ in  \eqref{actionprinciple} are used to constrain the velocities $\bu_i$ and $\widetilde{\bu}_e$ to the corresponding Eulerian coordinate so that
\begin{equation}\label{u-variables}
\bu_i(\bq,\bv,t)=\bv
\,,\ \qquad\
\widetilde{\bu}_e(\bq,\bc,t)=\bc
\,.
\end{equation}
{\color{black} Then, the coordinate $\bc$ is the electron fluctuation velocity, consistently with the velocity splitting $\boldsymbol{v}=\bc+\bV_e$ introduced in Section \ref{kin-model}.}
Notice that expanding all terms containing $\bu_i$ and $\widetilde{\bu}_e$ in \eqref{actionprinciple} yields the {\color{black}Euler-Poincar\'e form of Littlejohn's phase-space Lagrangian \cite{Littlejohn,SqQiTa}} for ions and electrons. This occurs because of the minus signs carried by the terms in $|\bu_i-\bv|^2$ and $|\widetilde{\bu}_e-\bc|^2$. These signs bring the present treatment {\color{black}(and the one in \cite{SqQiTa})} very close to that in \cite{CeHoHoMa},  although not identical. 

\subsection{The variations}
At this point, one needs to take variations. While this section gives an overview of the method, a more detailed discussion is found in Appendix \ref{App}. Following \cite{CeHoHoMa}, variations of the Lagrangian labels imply the following Eulerian relations:
\begin{align}\label{phasespacevariations}
\delta\bX_{i}&=\partial_t\boldsymbol{Y}_i+[{\color{black}\bX_{i},\boldsymbol{Y}_i}]
\,,\ \qquad\
\delta\widetilde{\bX}_{e}=\partial_t\widetilde{\boldsymbol{Y}}_e+[{\color{black}\widetilde{\bX}_{e},\widetilde{\boldsymbol{Y}}_e}]+[\bX_{{\bV}_e},\widetilde{\boldsymbol{Y}}_e]-[\bX_{\mathbf{W}},\widetilde{\bX}_{e}]
\,,
\\\label{phasespacevariations2}
\delta f_i&=-\nabla\cdot(f_i\,\boldsymbol{Y}_i)
\,,\ \qquad\qquad\,
\delta \tf_e=-\nabla\cdot(\tf_e\,\widetilde{\boldsymbol{Y}}_e)-\nabla\cdot(\tf_e\,\bX_{\mathbf{W}})
\,,
\end{align}
where $[\mathbf{P},\mathbf{R}]=(\mathbf{P}\cdot\nabla)\mathbf{R}-(\mathbf{R}\cdot\nabla)\mathbf{P}$ is the vector field commutator, while $\boldsymbol{Y}_i(\bq,\bv,t)$, $\widetilde{\boldsymbol{Y}}_e(\bq,\bc,t)$ and $\mathbf{W}(\bq,t)$ are arbitrary vector fields vanishing at $t_1$ and $t_2$. {\color{black}In particular, while the vector field $\mathbf{W}$ encodes variations of the electron mean flow, the phase-space vector field $\widetilde{\boldsymbol{Y}}_e$ comprises variations of the flow of electron fluctuations.} Here the vector field $\bX_{{\bV}_e}$ is constructed from $\bV_e$ as
$\bX_{{\bV}_e}=(\bV_e,\,\bc\cdot\nabla_{\!\bq}\bV_e)$ and analogously for $\bX_\bW$. Notice that, upon introducing the absolute velocity $\boldsymbol{v}:=\bc+\bV_e$, we may write by a slight abuse of ntation
\[
(\dot{\mathbf{x}},\dot{\boldsymbol{v}})_e=(\widetilde{\bu}_e,\,\widetilde{\mathbf{a}}_e)+(\bV_e,\,\bc\cdot\nabla\bV_e)=\widetilde{\bX}_{e}+\bX_{{\bV}_e}
\]
 The physical interpretation for the emergence of these expressions goes back to the fact that $\widetilde{\bX}_{e}$ is a phase-space vector field that is expressed in the frame moving with $\bV_e$; see also \cite{HoTr2011}, where the same method is applied to hybrid MHD models. Consequently, once variations \eqref{phasespacevariations} are taken in \eqref{actionprinciple}, expanding the terms in $\delta\ba_i$ and $\delta\tilde{\ba}_e$ yields \eqref{u-variables}.

In the next stage, following \cite{IlLa}, arbitrary variations $(\delta\varphi,\delta\bA)$ yield the first two equations in \eqref{kinetic-quantities}, i.e. Amp\`ere's law and charge neutrality
\begin{equation}\label{Ampere+Gauss}
{\color{black}\mu_0^{-1}}\nabla\times\bB={\color{black}q_e}n_e\bV_e+{\color{black}q_i}\!\int\!\bv f_i\,\de^3\bv
\,,\qquad\text{\ and\ }\qquad
{\color{black}q_e}n_e=-{\color{black}q_i}\!\int\! f_i\,\de^3\bv
\end{equation}
On the other hand, variations of $\bV_e$ are given by
\begin{equation}\label{varVe}
\delta\bV_e=\partial_t\bW+[\bV_e,\bW]
\end{equation}
At this point, one makes the crucial \emph{assumption of inertialess electrons} in the mean velocity equation: this step amounts exactly to letting 
\[
\epsilon\to0
,
\]
in the action functional \eqref{actionprinciple}, so that the electron mean flow contributions to the kinetic energy is neglected, while retaining its current contribution in Amp\`ere's law \eqref{Ampere+Gauss}. The limit $\epsilon\to0$ in the action \eqref{actionprinciple} yields the following expression for the electric field $\bE$
\begin{equation}\label{electricfield}
\bE=-\left(\frac{\partial\bA}{\partial t}+\nabla\varphi\right)
=-{\bV_e}\times\bB
+\frac1{q_e n_e}\nabla\cdot\widetilde{\Bbb{P}}_e
\,,
\end{equation}
Ordinarily, a choice of gauge would be necessary  to specify the dynamics of $\bA$. For example, in \cite{IlLa} the hydrodynamic gauge $\varphi+\bV_e\cdot\bA=0$  is chosen. However, taking the curl of the above equation yields \eqref{magneticfield}. 

Then, expanding the terms in $\delta\bu_i$ and $\delta\tilde{\bu}_e$ in \eqref{actionprinciple} yields the expressions of the accelerations $\ba_i$ and $\tilde{\ba}_e$. Eventually, one finds
\begin{align}\label{phasespace1}
&\bX_{i}=\left(\bv\,,\,\frac{q_i}{m_i}\left[\big(\bv-\bV_e\big)\times\bB+\frac1{q_e n_e}\nabla\cdot\widetilde{\Bbb{P}}_e\right]\right)
\\
\label{phasespace2}
&\widetilde{\bX}_{e}=\left(\bc\,,\,\bc\times\!\left(\frac{q_e}{m_e}\bB-\nabla\times\bV_e\right)-2\bc\cdot\nabla\bV_e+\frac1{m_e n_e}\nabla\cdot\widetilde{\Bbb{P}}_e\right)
.
\end{align}
Finally, following \cite{HoMaRa1998,CeHoHoMa,IlLa}, one recalls that \eqref{actionprinciple} can be expressed in terms of Lagrangian labels and derives the kinetic equations
\[
\frac{\partial f_i}{\partial t}+\nabla\cdot(f_i\,{\bX_i})=0
\,,\qquad
\frac{\partial \tf_e}{\partial t}+\nabla\cdot(\tf_e\,\widetilde{\bX}_e)+\nabla\cdot(\tf_e\,{\bX}_{\bV_e})=0
\]
by taking the time derivative of the Lagrange-to-Euler map for the two probability densities. Then, using the expressions \eqref{phasespace1}-\eqref{phasespace2}, one obtains the kinetic equations \eqref{electron-f} and \eqref{ion-f}.

{\color{black}

\subsection{Inertia of the electron mean flow\label{MeanInertia}}
As it emerges from the previous Section, the insertion of the parameter $\epsilon$ provides a formal mechanism for switching off the inertia of the electron mean flow. Notice that this is not the same as neglecting all electron inertial effects. This can be seen upon considering the electron velocity splitting $\boldsymbol{v}=\bc+\bV_e$. Indeed, upon introducing the convenient parameter $\epsilon$, the velocity splitting may be used to define a modified kinetic momentum  as $\mathbf{p}_\epsilon=m_e\bc+\epsilon m_e \bV_e$. It is clear that setting $\epsilon=1$ amounts to considering all electron inertial effects. For example, setting $\epsilon=1$ in \eqref{actionprinciple} amounts to including the electron inertial terms $m_eq_e^{-1}(\partial_t\bV_e+(\bV_e\cdot\nabla)\bV_e)$ in the right hand side of the second equality in \eqref{electricfield}. On the other hand, letting $\epsilon\to0$ neglects only those inertial effects arising from the electron mean flow (with velocity $\bV_e$), without imposing any assumption on the fluctuations. Consequently, although the inertia of the electron mean flow is discarded, the pressure term $({q_e n_e})^{-1}\nabla\cdot\widetilde{\Bbb{P}}_e$ is retained as it arises from inertial effects of the fluctuation dynamics with velocity $\bc$. This reflects in the fact that the electron mass still occurs in the model, although terms like $m_e\bV_e$ are all neglected.
}

\section{Electron pressure dynamics\label{elpressure}}

\subsection{Hybrid model with frozen-in electron pressure}

From the electron kinetics, one derives the evolution of the electron pressure tensor{\color{black}. This is found by taking the second-order moment of \eqref{electron-f}}, which reads
\begin{multline*}
\frac{\partial\widetilde{\Bbb{P}}_e}{\partial t}+(\bV_e\cdot\nabla)\widetilde{\Bbb{P}}+(\nabla\cdot\bV_e)\widetilde{\Bbb{P}}_e+\widetilde{\Bbb{P}}_e\cdot\nabla\bV_e+\big(\widetilde{\Bbb{P}}_e\cdot\nabla\bV_e\big)^T
\\
+\left(\frac{q_e}{m_e}\bB-\boldsymbol\omega_e\right)\!\times\widetilde{\Bbb{P}}_e-\widetilde{\Bbb{P}}_e\times\!\left(\frac{q_e}{m_e}\bB-\boldsymbol\omega_e\right)
=-\nabla\cdot\boldsymbol{\mathsf{Q}}_e
\,,
\end{multline*}
where $\boldsymbol{\mathsf{Q}}_e=\int\!\bc\bc\bc\,\tf_e\,\de^3\bc$ is the heat flux tensor.

According to the above equation, the main problem concerning fully kinetic models for collisionless reconnection  is that solving for two kinetic equations (ions and electrons) presents outstanding computational difficulties \cite{MaLaRi} arising from the emergence of infinite moment hierarchies. This has led various authors to  formulations of hybrid models, in which electron dynamics could be described by some kind of moment closure that retains the pressure tensor dynamics \cite{DiMaLaSeErBi}.  In particular, see   \cite{WiYiOmKaQu,YiWiGaBi,KuHeWi,KuHeBi,YiWi} and following papers by the same authors. In these works, the evolution of the electron pressure tensor dynamics is obtained by dropping the $\boldsymbol\omega_e$- and $\boldsymbol{\mathsf{Q}}_e$-terms  and upon replacing the $\bB$-terms {\color{black}by a relaxation term} as follows:
\begin{equation}\label{pressure-winske}
\frac{\partial\widetilde{\Bbb{P}}_e}{\partial t}+(\bV_e\cdot\nabla)\widetilde{\Bbb{P}}_e+(\nabla\cdot\bV_e)\widetilde{\Bbb{P}}_e+\widetilde{\Bbb{P}}_e\cdot\nabla\bV_e+\big(\widetilde{\Bbb{P}}_e\cdot\nabla\bV_e\big)^T=
-
 \frac{q_e|\bB|}{m_e\tau}\left(\widetilde{\Bbb{P}}_e-\mathsf{p}_e\mathbb{I}\right)
,
\end{equation}
where $\mathsf{p}_e$ is the scalar electron pressure and $\mathbb{I}$ is the identity matrix, while $\tau$ is a phenomenological time-scale introduced \emph{ad hoc} to match the available data.
Increasing values of $\tau$ yield higher reconnected flux, althuogh at some point singularities start to develop \cite{YiWiGaBi}. The above evolution equation yields reasonable results in many situations, although its underlying fundamental properties are not completely understood. Moreover, neglecting the heat flux tensor term may lead to physical inconsistencies \cite{KuHeWi}, while a detailed study of its contribution is found in \cite{KuHeBi}. Still, most hybrid models for collisionless reconnection tend to disregard heat flux contributions.

If we accept that the heat flux can be neglected $\nabla\cdot\boldsymbol{\mathsf{Q}}_e\simeq0$, then, the model \eqref{electron-f}-\eqref{kinetic-quantities} (which retains the Coriolis force terms $\boldsymbol\omega_e\times\widetilde{\Bbb{P}}_e-\widetilde{\Bbb{P}}_e\times\boldsymbol\omega_e$) yields the electron pressure dynamics in \eqref{pressure-eq}. Moreover, {\color{black} as pointed out in Section \ref{pressure}, the strong electron magnetization limit provides an interesting case of study. In this limit, the electron Larmor period $\omega_L^{-1}$ is much shorter than any other characteristic time scales in the system (i.e. $\omega/\omega_L<<1$, for any typical frequency $\omega$), and the electron Larmor radius $r_L$ much smaller than any other characteristic lengths (i.e. $r_L/\lambda<<1$, for any typical wavelength $\lambda$). This leads to the relation $\bB\times\widetilde{\Bbb{P}}_e-\widetilde{\Bbb{P}}_e\times\bB\simeq0$, so that} the electron pressure tensor becomes \emph{frozen} into the electron mean velocity {\color{black}or, equivalently,} the electron velocity correlations are constant along the electron mean flow. When this happens,  one can simply replace the kinetic equation \eqref{electron-f} by \eqref{pressure-eq}. This yields a hybrid model in which ion kinetics \eqref{ion-f} is coupled to an electron velocity $\bV_e$ transporting its own pressure tensor $\widetilde{\Bbb{P}}_e$, along the lines of  \cite{WiYiOmKaQu,YiWiGaBi,KuHeWi,KuHeBi,YiWi}. More explicitly, this procedure yields
\begin{align}\label{electron-pressure}
&
\frac{\partial\widetilde{\Bbb{P}}_e}{\partial t}+(\bV_e\cdot\nabla)\widetilde{\Bbb{P}}_e+(\nabla\cdot\bV_e)\widetilde{\Bbb{P}}_e
{\color{black}+\widetilde{\Bbb{P}}_e\times\boldsymbol\omega_e
-\boldsymbol\omega_e\times\widetilde{\Bbb{P}}_e}
+\widetilde{\Bbb{P}}_e\cdot\nabla\bV_e+\big(\widetilde{\Bbb{P}}_e\cdot\nabla\bV_e\big)^T
=0
\\\label{ion-f2}
&
\frac{\partial f_i}{\partial t}+\bv\cdot\frac{\partial f_i}{\partial \bq}+\frac{q_i}{m_i}\left[\big(\bv-\bV_e\big)\times\bB-{\color{black}\frac1{q_in_i}}\nabla\cdot\widetilde{\Bbb{P}}_e\right]\cdot\frac{\partial f_i}{\partial \bv}=0
\\\label{magneticfield2}
&
\frac{\partial\bB}{\partial t}
=\nabla\times\!\left({\bV_e}\times\bB
+{\color{black}\frac1{q_i n_i}}\nabla\cdot\widetilde{\Bbb{P}}_e
\right)
\end{align}
where 
\begin{align}
&\label{kinetic-quantities2}
n_i=\int\! f_i\,\de^3\bv
\,,\qquad\quad
q_in_i\bV_e=-{\color{black}\mu_0^{-1}}\nabla\times\bB+q_i\!\int\!\bv\, f_i\,\de^3\bv
\,.
\end{align}
%\noindent\comment{Add invariant traces}\noindent
\rem{ %%%%%%%%%%%%%%%%%%%%%%%%%%%%%%%%%
\begin{remark}[Strong electron magnetization limit]
The {\it strong electron magnetization limit} \cite{YiWi}
\[
\bB\times\widetilde{\Bbb{P}}_e-\widetilde{\Bbb{P}}_e\times\bB\simeq0
\]
is recovered in the present approach by neglecting the minimal coupling term
\[
q_e\!\int\!\bA\cdot\widetilde{\mathbf{u}}_e\,\de^3\bq\,\de^3\bc
\]
in the action \eqref{actionprinciple}, which then assumes that fluctuations are not directly coupled to the magnetic field.
\end{remark}
} %%%%%%%%%%%%%%%%%%%%%%%%%%%%%%%%%
{\color{black}It is perhaps worth emphasizing that the electron pressure and the magnetic field are consistently affected by the ion mean flow, as it emerges from Amp\`ere's law  in \eqref{kinetic-quantities2}.} 

We point out that the Coriolis force terms $\boldsymbol\omega_e\times\widetilde{\Bbb{P}}_e-\widetilde{\Bbb{P}}_e\times\boldsymbol\omega_e$ are strictly necessary for a frozen-in electron pressure. Indeed, dropping these terms completely breaks this frozen-in condition. On the other hand, we should emphasize that this result is obtained by neglecting the heat flux contribution, which may not always be justifiable. The available data depend mainly on the particular situation that is being considered each time and no general statement is available in this regard. 

{\color{black}
\begin{remark}[Models with scalar electron pressure]{\color{black}
It is well known that a purely scalar electron pressure is not sufficient to generate reconnection. On the other hand, a scalar electron pressure is still used in many situations involving the generation of energetic particles. Examples are given by Field Reversed Configuration devices and certain space plasmas \cite{PaBeFuTaStSu,WiYiOmKaQu}. In these contexts, hybrid models can be formulated, as described in \cite{WiYiOmKaQu}.
When the kinetic features of electron dynamics are neglected and one is interested only in the quantities $\bV_e$ and $n_e$, one only needs to replace the electron pressure tensor contribution $\nabla\cdot\widetilde{\Bbb{P}}_e$ in \eqref{ion-f}-\eqref{magneticfield} by the pressure  gradient $\nabla\mathsf{p}_e$, as it is given in terms of the electron density $n_e$ by an appropriate equation of state. Upon recalling charge neutrality and Amp\`ere's law in \eqref{Ampere+Gauss}, we find the resulting model
\[
\frac{\partial f_i}{\partial t}+\bv\cdot\frac{\partial f_i}{\partial \bq}+\frac{q_i}{m_i}\left[\big(\bv-\bV_e\big)\times\bB+\frac1{q_e n_e}\nabla\mathsf{p}_e\right]\cdot\frac{\partial f_i}{\partial \bv}=0
\,,\qquad\quad
\frac{\partial\bB}{\partial t}
+\nabla\times(\bB\times\bV_e)
=0
\,,
\]
which was also analyzed in \cite{Tronci2010} from the Hamiltonian viewpoint. Here, the scalar pressure arises from an internal energy function $e(n_e)$ as $\mathsf{p}_e:=n_e^2 e'(n_e)$, where prime denotes total derivative.
}
\end{remark}

\subsection{Extended Hall magnetohydrodynamics\label{extendedHallMHD}}
As it is well known, ideal Hall MHD does not allow for magnetic reconnection. Indeed, rather than freezing magnetic field lines along the ion mean flow (as in ideal MHD), Hall MHD freezes the magnetic field along the electron fluid velocity. On the other hand, in collisionless kinetic reconnection, the magnetic frozen-in condition is broken by both the electron pressure tensor and the inertial terms generated by the electron mean flow (when they are relevant). Still, one would prefer to deal with fluid models because of their various numerical advantages over fully kinetic simulations. In this light, in \cite{YiWi} a comparison study was presented between fully kinetic simulations and an extended version of ideal Hall MHD, which is modified to account for effects of the electron pressure tensor (thereby allowing for magnetic reconnection). This extended Hall MHD model arises by  coupling the fluid closure of ion kinetics \eqref{ion-f} with the pressure tensor dynamics in \eqref{pressure-winske}. 

At this point, one is naturally led to ask what type of extended Hall MHD is provided by the model \eqref{electron-f}-\eqref{kinetic-quantities} presented here.  In the strong electron magnetization limit, the answer is easily found by taking the fluid closure of  the ion kinetic equation \eqref{ion-f2} in the system \eqref{electron-pressure}-\eqref{magneticfield2}, which then becomes
\begin{align}\label{electron-pressure-bis}
&
\frac{\partial\widetilde{\Bbb{P}}_e}{\partial t}+(\bV_e\cdot\nabla)\widetilde{\Bbb{P}}_e+(\nabla\cdot\bV_e)\widetilde{\Bbb{P}}_e
+\widetilde{\Bbb{P}}_e\times\boldsymbol\omega_e
-\boldsymbol\omega_e\times\widetilde{\Bbb{P}}_e
+
\widetilde{\Bbb{P}}_e\cdot\nabla\bV_e+\big(\widetilde{\Bbb{P}}_e\cdot\nabla\bV_e\big)^T
=0
\\\label{ion-f2-bis}
&
m_i n_i\!\left(\frac{\partial \bV_i}{\partial t}+\bV_i\cdot\nabla\bV_i\right)\!
=
-\nabla\mathsf{p}_i-\nabla\cdot\widetilde{\Bbb{P}}_e-\mu_0^{-1}\bB\times\nabla\times\bB
\,,\qquad\!
\frac{\partial n_i}{\partial t}+\nabla\cdot(n_i\bV_i)=0
\\\label{magneticfield2-bis}
&
\frac{\partial\bB}{\partial t}
=\nabla\times\!\left({\bV_e}\times\bB+
\frac1{q_i n_i}\nabla\cdot\widetilde{\Bbb{P}}_e
\right)
,
\end{align}
together with Amp\`ere's law $q_in_i\bV_e=-{\color{black}\mu_0^{-1}}\nabla\times\bB+q_i n_i\bV_i$. Then, it is not surprising that the only differences between the extended Hall MHD in \cite{YiWi} and the one emerging from \eqref{electron-f}-\eqref{kinetic-quantities} lies in the electron pressure dynamics, since \eqref{electron-pressure-bis} includes the Coriolis-force terms $\widetilde{\Bbb{P}}_e\times\boldsymbol\omega_e
-\boldsymbol\omega_e\times\widetilde{\Bbb{P}}_e$ that are absent in \eqref{pressure-winske}. Being a fluid closure of \eqref{electron-f}-\eqref{kinetic-quantities}, the system \eqref{electron-pressure-bis}-\eqref{magneticfield2-bis} provides a simple and  interesting  case  for a detailed study of the physical properties of the proposed new model, in comparison to previous models. This will be the subject of future work.
}

\section{Stationary equations and constants of motion\label{morestuff-sec}}

In this Section, we shall discuss a few properties of the stationary solutions of the model \eqref{electron-f}-\eqref{kinetic-quantities}. While a nonlinear stability analysis requires knowledge of constants of motion that are not available a priori, some insight can be provided by the linearized equations. However, given the level of difficulty of the latter, this topic is left for further research in this direction. In this Section, we shall make some comments on how the system \eqref{electron-f}-\eqref{kinetic-quantities} possesses Harris' current sheet solution and we shall show how the available constants of motion are insufficient for studying equilibria with non-zero current. 

\subsection{Stationary solutions and Harris' current sheets}

{\color{black}This Section proves that the proposed model \eqref{electron-f}-\eqref{kinetic-quantities} admits the Harris' current sheet configuration as an equilibrium solution.} Let us start our discussion by writing the stationary equations corresponding to \eqref{electron-f}-\eqref{kinetic-quantities}. One has
\begin{align}\label{electron-f-eq}
&
\big(\bc+\bV_e\big)\cdot\frac{\partial\tf_e}{\partial \bq}+\left[\bc\times\!\left(\frac{q_e}{m_e}\bB-\nabla\times\bV_e\right)-\bc\cdot\nabla\bV_e+\frac1{m_e n_e}\nabla\cdot\widetilde{\Bbb{P}}_e\right]\cdot\frac{\partial\tf_e}{\partial \bc}=0
\\\label{ion-f-eq}
&
\bv\cdot\frac{\partial f_i}{\partial \bq}+\frac{q_i}{m_i}\left[\big(\bv-\bV_e\big)\times\bB+\frac1{q_e n_e}\nabla\cdot\widetilde{\Bbb{P}}_e\right]\cdot\frac{\partial f_i}{\partial \bv}=0
\\\label{magneticfield-eq}
&
\nabla\times\!\left({\bV_e}\times\bB
-\frac1{q_en_e}\nabla\cdot\widetilde{\Bbb{P}}_e
\right)=0
\end{align}
with 
\begin{align}
&\label{kinetic-quantities-eq}
q_en_e=-q_i\!\int\! f_i\,\de^3\bv
\,,\qquad
q_en_e\bV_e={\color{black}\mu_0^{-1}}\nabla\times\bB-q_i\!\int\!\bv\, f_i\,\de^3\bv
\,.
\end{align}
Now, following Harris' work \cite{Harris}, we look for stationary states such that $\bE=0$. Then, upon recalling \eqref{electricfield}, we have
\[
{\bV_e}\times\bB
=\frac1{q_e n_e}\nabla\cdot\widetilde{\Bbb{P}}_e
\,.
\]
{\color{black}At this point,} we observe that the electron {\color{black}equilibrium} solution has the general form
\begin{equation}\label{equilibrium-f}
\tf_e=\tf_e\!\left(\frac12\left|\bc+\bV_e\right|^2-\frac12\left|\bV_e\right|^2\right)
=f_e\!\left(\frac12\left|\boldsymbol{v}\right|^2-\frac12\left|\bV_e\right|^2\right)
\end{equation}
where we introduce $\boldsymbol{v}:=\bc+\bV_e$. {\color{black}The  solution above means that, at the  equilibrium, the electron distribution depends on the difference between the total kinetic energy $m_ev^2/2$ and the kinetic energy of the mean flow $m_e V_e^2/2$, which is consistent with the initial assumption that neglects the inertial effects of electron mean flow (with velocity $\bV_e$).
}
The equilibrium solution \eqref{equilibrium-f} can be {\color{black}easily} verified by replacing {\color{black}its} expression into equation \eqref{electron-f-eq}, which may be written in terms of the canonical Poisson bracket $\{\cdot,\,\cdot\}$ as
\[
\left\{\tf_e,\frac12\left|\bc+\bV_e\right|^2-\frac12\left|\bV_e\right|^2\right\}+\frac{q_e}{m_e}\big(\bc+\bV_e\big)\times\bB\cdot\frac{\partial\tf_e}{\partial \bc}=0
\,.
\]
{\color{black}In order to proceed further, it is convenient to express \eqref{electron-f-eq} in terms of the the  electron velocity coordinate $\boldsymbol{v}$. Then, upon using \eqref{kinetic-quantities-eq},  the Vlasov densities $f_e(\bq,\boldsymbol{v})$ and $f_i(\bq,\bv)$ are shown to satisfy}
\begin{align}\label{electron-f-eq2}
&
\boldsymbol{v}\cdot\frac{\partial f_e}{\partial \bq}+\left(\frac{q_e}{m_e}\boldsymbol{v}\times\bB-\boldsymbol{v}\times\nabla\times\bV_e+\nabla\bV_e\cdot\bV_e\right)\cdot\frac{\partial f_e}{\partial \boldsymbol{v}}=0
\\\label{ion-f-eq2}
&
\bv\cdot\frac{\partial f_i}{\partial \bq}+\frac{q_i}{m_i}\bv\times\bB\cdot\frac{\partial f_i}{\partial \bv}=0
\,,
\end{align}
where the terms in $\bV_e$ {\color{black}emerge in the electron equation as another consequence of} the assumption of negligible inertia of the mean flow. {\color{black}(Observe that expressing electron kinetics in terms of the total velocity coordinate $\boldsymbol{v}$ produces in \eqref{electron-f-eq2} the term $\nabla\bV_e\cdot\bV_e$, which was absent in \eqref{electron-f-eq}).}

{\color{black}Finally, we follow \cite{Harris} and we} restrict to the case when $\bV_e$ and $\bV_i$ are spatially constant and $\bV_e=-\bV_i$. {\color{black}This step reduces the problem \eqref{electron-f-eq2}-\eqref{ion-f-eq2} to finding the equilibrium solutions of} the usual  Vlasov equations
\begin{align}\label{electron-f-eq3}
&
\boldsymbol{v}\cdot\frac{\partial f_e}{\partial \bq}+\frac{q_e}{m_e}\boldsymbol{v}\times\bB\cdot\frac{\partial f_e}{\partial \boldsymbol{v}}=0
\\\label{ion-f-eq3}
&
\bv\cdot\frac{\partial f_i}{\partial \bq}+\frac{q_i}{m_i}\bv\times\bB\cdot\frac{\partial f_i}{\partial \bv}=0
\,,
\end{align}
{\color{black}which then possess the Harris current sheet solution in \cite{Harris},} with $q_e=e=-q_i$ and
\[
n_e=\!\int\! f_i\,\de^3\bv
\,,\qquad
2n_e\bV_e={\color{black}\mu_0^{-1}}\nabla\times\bB
\,.
\]
{\color{black}This shows that the proposed model \eqref{electron-f}-\eqref{kinetic-quantities} comprises the Harris current sheet as an equilibrium configuration.}

\begin{remark}[Static equilibria]
Notice that for static equilibria $\bV_e=\int\!f_i\,\bv\,\de^3\bv=0$, one has
\begin{align}\label{electron-f-eq-0}
&
\bc\cdot\frac{\partial\tf_e}{\partial \bq}+\left[\frac{q_e}{m_e}\bc\times\bB+\frac1{q_e n_e}\nabla\cdot\widetilde{\Bbb{P}}_e\right]\cdot\frac{\partial\tf_e}{\partial \bc}=0
\\\label{ion-f-eq-0}
&
\bv\cdot\frac{\partial f_i}{\partial \bq}+\frac{q_i}{m_i}\left[\bv\times\bB+\frac1{q_e n_e}\nabla\cdot\widetilde{\Bbb{P}}_e\right]\cdot\frac{\partial f_i}{\partial \bv}=0
\\\label{magneticfield-eq-0}
&
\nabla\times\!\left(
\frac1{n_e}\nabla\cdot\widetilde{\Bbb{P}}_e
\right)=0
\end{align}
and if we assume $\bE=(q_e n_e)^{-1\,}\nabla\cdot\widetilde{\Bbb{P}}_e=0$,  we have 
\begin{align}\label{electron-f-eq-01}
&
\bc\cdot\frac{\partial\tf_e}{\partial \bq}+\frac{q_e}{m_e}\bc\times\bB\cdot\frac{\partial\tf_e}{\partial \bc}=0
\\\label{ion-f-eq-01}
&
\bv\cdot\frac{\partial f_i}{\partial \bq}+\frac{q_i}{m_i}\bv\times\bB\cdot\frac{\partial f_i}{\partial \bv}=0
\end{align}
along with the zero-current relation $\nabla\times\bB=0$. Here, equation \eqref{electron-f-eq-01} is the same as  \eqref{electron-f-eq3} since $\bV_e=0$. {\color{black} In this configuration, both ions and electrons are dominated only by fluctuation dynamics.}
\end{remark}

\subsection{Total energy and constants of motion}

In the search for new physical models, it is essential that the energy is an exact constant of motion {\color{black}(when dissipation effects are neglected)}. For example, several models that are derived by making assumptions directly in the equations of motion do not possess this vital property (see \cite{Tronci2010} for a similar discussion on hybrid MHD models) {\color{black}and energy conservation holds only under certain approximations}. In the case of system \eqref{electron-f}-\eqref{kinetic-quantities}, exact conservation of energy is guaranteed by the underlying variational principle \eqref{actionprinciple}. When this happens, the system is Hamiltonian, which means it conserves energy {\color{black}exactly} and it possesses a Poisson bracket structure. The latter property can be verified explicitly also in the present context, although this requires cumbersome calculations, which in turn do not add anything to the physical content of equations \eqref{electron-f}-\eqref{kinetic-quantities}. Therefore the explicit identification of the Poisson bracket structure underlying  \eqref{electron-f}-\eqref{kinetic-quantities} is left for future developments. {\color{black}In the simpler case when electrons are treated as an ordinary fluid, the Hamiltonian structure of the resulting  equations was presented in \cite{Tronci2010}. Moreover, treating both ions and electrons as fluid components yields ideal Hall MHD, whose Hamiltonian structure was presented in \cite{Holm1987}, along with the Lyapunov stability analysis.} Here, we focus only on the constants of motion.

{\color{black}As it can be easily verified by an explicit calculation}, the following expression of the total energy remains constant under the dynamics \eqref{electron-f}-\eqref{kinetic-quantities}:
\begin{equation}
\mathcal{E}(f_i,\tf_e,\bA)=\frac{m_i}2\int\!f_i\,|\bv|^2\,\de^3\bq\,\de^3\bv+\frac{m_e}2\int\!\tf_e\,|\bc|^2\,\de^3\bq\,\de^3\bc+\frac1{2{\color{black}\mu_0}}\int\!\left|\nabla\times\bA\right|^2\de^3\bq
\,.
\end{equation}
{\color{black}This is given by the sum of ion and electron kinetic energies, plus the magnetic energy (last term above).
Notice that the electron kinetic energy (second term above) differs from the usual expression} because only the fluctuation velocity $\bc=\boldsymbol{v}-\bV_e$ {\color{black}is involved}.

Other constants of motion are present in the dynamics \eqref{electron-f}-\eqref{kinetic-quantities}. Actually, these are  two separate families of constants that are defined as follows:
\[
C_i=\int\!\Phi_i(f_i)\,\de^3\bq\,\de^3\bv
\,,\qquad
C_e=\int\!\Phi_e(\tf_e)\,\de^3\bq\,\de^3\bv
\]
where $\Phi_i$ and $\Phi_e$ are arbitrary functions of their arguments. {\color{black} These constants should come as no surprise, since they are the usual quantities that are conserved by Vlasov dynamics. In the simplest case when $\Phi_i$ and $\Phi_e$ are the identity, the above constants return conservation of the total number of particles.}

\begin{remark}[Energy-Casimir method for nonlinear stability]

Following \cite{HoMaRaWe}, one can use the constants of motion to perform a nonlinear stability analysis for the system \eqref{electron-f}-\eqref{kinetic-quantities}. In particular, this method identifies the equilibria by setting
\begin{equation}\label{stability1}
\delta\!\left(\mathcal{E}+C_i+C_e\right)=0
\end{equation}
and gives Lyapunov stability whenever
\begin{equation}\label{stability2}
\delta^2\!\left(\mathcal{E}+C_i+C_e\right)>0
\end{equation}
However, the success of this method depends on the number of constants of motion that are available. It turns out that the three constants of motion found in this Section restrict to consider only equilibria with zero current, thereby eliminating Harris' solution from the treatment. Indeed, \eqref{stability1} yields $\nabla\times\bB=0$ at the equilibrium. At this point, one can try to find new constants of motion to enrich the stability analysis. On the other hand, this task can be difficult to reach and it requires finding the explicit Poisson bracket for the system \eqref{electron-f}-\eqref{kinetic-quantities}. This will be subject of future work.
\end{remark}

\bigskip

\section{Conclusions}

This paper has presented the new fully kinetic model \eqref{electron-f}-\eqref{kinetic-quantities} for collisionless reconnection, whose main novelty is the introduction of a Coriolis force term 
\[
-\,\bc\times\nabla\times\bV_e\cdot\frac{\partial\tf_e}{\partial \bc}
\]
in the electron kinetic equation. Unlike previous models, this term was obtained by neglecting the inertia of the electron mean flow in the variational principle underlying the Maxwell-Vlasov system. Perhaps non surprisingly, the Coriolis force arises from a change of reference, similarly to the results in \cite{ThMc}. 

The principal consequence of introducing the Coriolis force is that neglecting the heat flux contribution  yields a \emph{frozen-in condition} for the electron pressure tensor, in presence of a strong electron magnetization. In turn, this frozen-in condition can be used to formulate new hybrid models, along the lines of \cite{KuHeWi} and subsequent papers (analogue hybrid models were also presented in \cite{DiMaLaSeErBi}). Also, a new variant of extended Hall MHD becomes available in Section \ref{extendedHallMHD}. The frozen-in condition confers the electron pressure an intrinsic Lagrangian meaning, which may open various possibilities regarding simulation techniques.

The last part of the paper has presented general considerations about the stationary solutions of \eqref{electron-f}-\eqref{kinetic-quantities}. Various specializations were considered and it was shown how the stationary equations comprise Harris' solution of the Maxwell-Vlasov system. Moreover, the explicit expression of the conserved total energy was provided, along with two different families of constants of motion. The question of how constants of motion can be used to perform a nonlinear stability analysis is left for future studies. This requires further constants that hopefully can be found from the explicit form of the Poisson bracket underlying \eqref{electron-f}-\eqref{kinetic-quantities}. This Poisson bracket is the main object that underlies the Casimir method for Lyapunov stability and its identification can be be realized by cumbersome computations in the Hamiltonian framework.

\subsection*{Acknowledgements} The author is indebted with Darryl Holm, Giovanni Lapenta, Philip Morrison and Emanuele Tassi for many helpful discussions on these and related topics. {\color{black} Also, the author is very grateful to both referees for their extremely valuable comments and remarks, which significantly improved the exposition.}

\smallskip

{\color{black}

\appendix

\section{Explicit derivation of the model\label{App}}

Although the derivation of model \eqref{electron-f}-\eqref{kinetic-quantities} can be performed by simply using the variations \eqref{phasespacevariations}-\eqref{phasespacevariations2} and \eqref{varVe} directly in the variational principle \eqref{actionprinciple}, it is less cumbersome to approach this matter at a more general level. By using the general theory in \cite{HoTr2011} (see Section 4.2 therein), one finds that applying the variations \eqref{phasespacevariations}-\eqref{phasespacevariations2} and \eqref{varVe} to an action principle of the type
\[
\delta\int_{t_1}^{t_2}\!\ell(\bX_i,\widetilde{\bX}_e,\bV_e,f_i,\tf_e,\bA,\varphi)\,\de t=0
\]
yields the following equations of motion:
\begin{align}
&\frac{\partial}{\partial t}\frac{\delta \ell}{\delta \bX_i}+\pounds_{\bX_i}\frac{\delta \ell}{\delta \bX_i}=
f_i\nabla\frac{\delta \ell}{\delta f_i}
\label{Xi}
\\
&\frac{\partial}{\partial t}\frac{\delta \ell}{\delta \bV_e}+\pounds_{\bV_e}\frac{\delta\ell}{\delta \bV_e}
=
%\frac{\delta l}{\delta \tf_e}\diamond \tf_e
-
\int\!\left(\!\pounds_{\tX_e}\frac{\delta \ell}{\delta
\tX_e}-\tf_e\nabla\frac{\delta\ell}{\delta\tf_e}\right)_{\!\!\bq}
\!\de^3\bc
+
\frac{\partial}{\partial\mathbf{x}}\cdot\!\int\!\bc\!\left(\!\pounds_{\tX_e}\frac{\delta
\ell}{\delta \tX_e}-\tf_e\,\nabla\frac{\delta\ell}{\delta\tf_e}\right)_{\!\!\bc}
\!\de^3\bc
\label{Ve}
\\
&\frac{\partial}{\partial t}\frac{\delta \ell}{\delta \tX_e}+\pounds_{\tX_e+\bX_{\bV_e}}\frac{\delta \ell}{\delta \tX_e}=
 \tf_e\nabla\frac{\delta \ell}{\delta \tf_e}
\label{Xe}
\\
&\frac{\partial f_i}{\partial t}+\nabla\cdot(f_i{\bX_i})=0
\,,\qquad
\frac{\partial \tf_e}{\partial t}+\nabla\cdot\big(\tf_e\,\tX_e+\tf_e\bX_{\bV_e}\big)=0
\label{effes}
\,,
\end{align}
where $\pounds$ denotes Lie derivative (see below), while the notation $\left(\cdot\right)_{\bq}$ and $\left(\cdot\right)_{\bc}$ stands for projection on the spatial and velocity components (e.g. $(\nabla\phi)_{\bq}=\partial\phi/\partial\bq$ and $(\nabla\phi)_{\bc}=\partial\phi/\partial\bc$). 
These equations possess a deep geometric structure arising from their corresponding formulation in terms of Lagrangian variables. In particular,   the constrained variations \eqref{phasespacevariations}-\eqref{phasespacevariations2} and \eqref{varVe} for the Eulerian variables arise from arbitrary variations of their corresponding Lagrangian variables, see  \cite{HoTr2011}.

Then, arbitrary variations in the electromagnetic potentials $(\varphi,\bA)$ give also
\begin{equation}\label{Amper+Gauss2}
\frac{\delta \ell}{\delta \bA}=0
\,,\qquad\quad
\frac{\delta \ell}{\delta \varphi}=0
\,.
\end{equation}
Here we have used the following definition of functional derivative
\[
\delta \mathcal{F}(\boldsymbol\phi):=\int_{\mathcal{D}}\frac{\delta\mathcal{F}}{\delta \boldsymbol\phi}\cdot\delta\boldsymbol\phi
\,,
\]
where $\mathcal{F}$ is any functional of some function (possibly, a vector function) $\boldsymbol\phi$ on some domain $\mathcal{D}$ in either the configuration space or  the phase-space. Also, we made use of the Lie derivative operator \cite{CeHoHoMa,HoMaRa1998,HoTr2011,IlLa,SqQiTa}
\[
\pounds_{\bU}\frac{\delta\mathcal{F}}{\delta \bU}=(\bU\cdot\nabla)\frac{\delta\mathcal{F}}{\delta \bU}
+(\nabla\cdot\bU)\frac{\delta\mathcal{F}}{\delta \bU}
+
\nabla\bU\cdot\frac{\delta\mathcal{F}}{\delta \bU}
\,,
\]
which can be defined on either the configuration space or the phase-space. Here $\bU$ is some vector field and $\mathcal{F}$ is an arbitrary functional.

At this point, it suffices to compute the variational derivatives of the Lagrangian in \eqref{actionprinciple} and then to replace them in \eqref{Xi}-\eqref{Xe}. This process requires expanding the variational principle \eqref{actionprinciple}, whose Lagrangian can be written in the form
\begin{align}
\ell(\bX_i,\widetilde{\bX}_e,&\bV_e,f_i,\tf_e,\bA,\varphi)=
\nonumber
\\=&
\int\! f_i\,(m_i\bv+q_i\bA)\cdot\bu_i\,\de^3\bq\,\de^3\bv
-\frac{m_i}2\!\int \!f_i\,|\bv|^2\,\de^3\bq\,\de^3\bv
-q_i\!\int\!\varphi f_i\,\de^3\bv \,\de^3\bq
\nonumber
\\&
+\!\int \!\tf_e\,(m_e\bc+\epsilon m_e\bV_e+q_e\bA)\cdot\widetilde{\bu}_e\,\de^3\bq\,\de^3\bc
-\frac{m_e}2\!\int \!\tf_e\,|\bc|^2\,\de^3\bq\,\de^3\bc
-q_e\!\int\!\varphi \tf_e\,\de^3\bc\,\de^3\bq
\nonumber
\\&
+\epsilon^2\frac{m_e}2\!\int \!\tf_e\,|\bV_e|^2\,\de^3\bq\,\de^3\bc
+q_e\!\int\! \tf_e\bV_e\cdot\bA\,\de^3\bc\,\de^3\bq-\frac1{2\mu_0}\int|\nabla\times\bA|^2\,\de^3\bq
\,.
\label{actionprinciple2}
\end{align}
Then, we have
\begin{align}
\delta\ell=&\int\!\delta f_i\left[(m_i\bv+q_i\bA)\cdot\mathbf{u}_i-\frac{m_i}2|\bv|^2-q_i\varphi\right]\de^3\mathbf{x}\de^3\bv
\label{fi-var}
\\&
+
\int\!\delta\tf_e
\left[(m_i\bc+\epsilon m_e\bV_e+q_e\bA)\cdot\widetilde{\mathbf{u}}_e-\frac{m_e}2|\bc|^2+\epsilon^2\frac{m_e}2|\bV_e|^2+q_e\bV_e\cdot\bA-q_e\varphi\right]\de^3\mathbf{x}\de^3\bc
\label{fe-var}
\\&
+
\int\!
\delta\bA\cdot\left[q_i\int\!f_i\bu_i\,\de^3\bv+q_e\left(\int\!\tf_e\widetilde{\bu}_e\,\de^3\bc+\bV_e\int\!f_e\,\de^3\bc\right)-\mu_0^{-1}\nabla\times\bB\right]\de^3\mathbf{x}
\label{Avariation}
\\&
+
\int\!
\delta\bV_e\cdot\left[(\epsilon^2 m_e\bV_e+\bA)\int\!\tf_e\,\de^3\bc+\epsilon m_e\int\!\tf_e\widetilde{\bu}_e\,\de^3\bc
\right]\de^3\mathbf{x}
\label{Vvariation}
\\&
+
\int\!
\delta\bu_i\cdot\big[f_i(m_i\bv+q_i\bA)\big]\,\de^3\mathbf{x}\de^3\bv
+
\int\!
\delta\widetilde{\bu}_e\cdot\left[\tf_e(m_e\bc+\epsilon m_e\bV_e+q_e\bA)\right]\de^3\mathbf{x}\de^3\bc
\label{psvar}
\\&
+
\int\!
\delta\varphi\left[q_i\int\! f_i\,\de^3\bv+q_e\int\!\tf_e\,\de^3\bc\right]\de^3\mathbf{x}
\label{phivariation}
\end{align}
In order to obtain the explicit equations of motion, one starts by inserting the functional derivatives $\delta\ell/\delta\bX_i$ and $\delta\ell/\delta\widetilde{\bX}_e$ (arising from line \eqref{psvar}) in the relations \eqref{Xi} and \eqref{Xe}. Notice that the latter require the functional derivatives $\delta\ell/\delta f_i$ and $\delta\ell/\delta \tf_e$ arising from lines \eqref{fi-var} and \eqref{fe-var}. Then, upon recalling $\bX_i=(\bu_i,\ba_i)$ and $\widetilde{\bX}_e=(\widetilde{\bu}_e,\widetilde{\ba}_e)$, we notice that ${\delta\ell/\delta\ba_i}={\delta\ell/\delta\tilde{\ba}_e}=0$. Also, taking the $\ba_i$-components of \eqref{Xi} and the $\widetilde{\ba}_e$-components of \eqref{Xe} yields \eqref{u-variables}. It is useful to notice that 
replacing $\widetilde{\bu}_e=\bc$ in the second of \eqref{effes} gives the continuity equation $\partial_t n_e+\nabla\cdot(n_e\bV_e+\widetilde{\bK}_e)=0$, with $n_e=\int\!\tf_e\,\de^3\bc$ and $\widetilde{\bK}_e=\int\!\bc\,\tf_e\,\de^3\bc$ (notice that we do not impose $\widetilde{\bK}_e=0$ for the moment).

At this point, it is convenient to insert the functional derivative $\delta\ell/\delta\bV_e$ arising from line \eqref{Vvariation} in equation \eqref{Ve}. From now on, we let $\epsilon\to0$. Upon recalling the functional derivatives $\delta\ell/\delta f_i$ and $\delta\ell/\delta \tf_e$ arising from lines \eqref{fi-var} and \eqref{fe-var}, equation \eqref{Ve} gives
\[
q_e\left(\frac{\partial}{\partial t}+\pounds_{\bV_e}\right)( n_e\bA)
=
\int\!\tf_e\frac{\partial}{\partial\bq}\Big(q_e\bA\cdot\bV_e+q_e\bA\cdot\bc-q_e\varphi\Big)\,\de^3\bc
-
\frac{\partial}{\partial\bq}\cdot\int\!\tf_e\,\bc\left(m_e\bc+q_e\bA\right)\de^3\bc
\,,
\]
which becomes
\begin{equation}\label{elettr}
q_en_e\left(\frac{\partial\bA}{\partial t}+\nabla\varphi\right)
=q_en_e{\bV_e}\times\bB
-\nabla\cdot\widetilde{\Bbb{P}}_e+q_e\widetilde{\bK}_e\times\bB
\end{equation}
Then, taking the $\widetilde{\bu}_e$-components of \eqref{Xe} yields
\[
q_e\frac{\partial\bA}{\partial t}+\Big[q_e(\bc+\bV_e)\cdot\nabla\bA+m_e(\widetilde{\ba}_e+\bc\cdot\nabla\bV_e)\Big]
+
\nabla\bV_e\cdot\left(m_e\bc+q_e\bA\right)=q_e\nabla\bA\cdot\bc+q_e\nabla\!\left(\bA\cdot\bV_e+\varphi\right)
\,,
\]
which gives 
\[
\widetilde{\ba}_e+\bc\cdot\nabla\bV_e=\bc\times\!\left(\frac{q_e}{m_e}\bB-\nabla\times\bV_e\right)-\bc\cdot\nabla\bV_e+\frac1{m_e n_e}\nabla\cdot\widetilde{\Bbb{P}}_e-q_e\widetilde{\bK}_e\times\bB
\]
Notice that, upon setting $\widetilde{\bK}_e=0$, the relation above returns \eqref{phasespace2}. In order to prove that $\widetilde{\bK}_e=0$ is consistent with the resulting dynamics, we insert $\widetilde{\bu}_e=\bc$ (from equation \eqref{u-variables}) and the last relation above in the second of \eqref{effes} to obtain
\[
\frac{\partial\tf_e}{\partial t}+\big(\bc+\bV_e\big)\cdot\frac{\partial\tf_e}{\partial \bq}+\left[\frac{q_e}{m_e}\Big(\bE+(\bc+\bV_e)\times\!\bB\Big)-\bc\cdot\nabla\bV_e-\bc\times\nabla\times\bV_e\right]\cdot\frac{\partial\tf_e}{\partial \bc}=0
\,.
\]
Then, taking the first-order moment of the kinetic equation above yields the following dynamics for $\widetilde{\bK}_e=\int\!\bc\,\tf_e\,\de^3\bc$:
\[
\frac{\partial\widetilde{\bK}_e}{\partial t}+\pounds_{\bV_e}\widetilde{\bK}_e=0
\,,
\]
so that $\widetilde{\bK}_e=0$ is preserved by the dynamics if $\widetilde{\bK}_e$ vanishes at the initial time. Then, from now on we set $\widetilde{\bK}_e=0$ and we obtain \eqref{electron-f} and \eqref{magneticfield} (by taking the curl of \eqref{elettr}).
By proceeding analogously, one  takes the ${\bu}_i$-components of \eqref{Xi} to prove \eqref{phasespace1}, which in turn yields \eqref{ion-f}. Also, we see that inserting the functional derivatives $\delta\ell/\delta\bA$ and $\delta\ell/\delta\varphi$ (arising from lines \eqref{Avariation} and \eqref{phivariation}) in  \eqref{Amper+Gauss2} give Amp\`ere's law and charge neutrality in \eqref{Ampere+Gauss}, respectively. 

It can be interesting to observe that the Euler-Poincar\'e structure of the variations \eqref{phasespacevariations}-\eqref{phasespacevariations2} implies conservation of a circulation integral. This arises from a general statement for Euler-Poincar\'e variational principles  \cite{HoMaRa1998,CeHoHoMa}. For the case of the variational principle \eqref{actionprinciple}, equations \eqref{Xi} and \eqref{Xe} imply (respectively) the following Poincar\'e relative integral invariants \cite{Arnoldbook}:
\[
\frac{\de}{\de t}\oint_{\eta_t(\bX_i)}(m_e\bv+q_e\bA)\cdot\de\mathbf{x}=0
\,,\qquad\quad 
\frac{\de}{\de t}\oint_{\gamma_t(\bX_e)}(m_e\bc+q_e\bA)\cdot\de\mathbf{x}=0
\]
where $\eta_t(\bX_i)$ and $\gamma_t(\bX_e)$ are two arbitrary time-dependent loops in phase-space, moving with the vector fields $\bX_i$ and $\bX_e=\widetilde{\bX}_{e}+\bX_{{\bV}_e}$, respectively. Analogous relations were found also in the context of the hybrid kinetic-MHD models in \cite{HoTr2011} (see Sections 4.3 and 5.3 therein). Notice that the result obtained for electron dynamics (second relation above) differs from that for ion kinetics (first relation above), which is a standard result in classical mechanics \cite{Arnoldbook}. Indeed, the integrand in the second relation  is not the total canonical momentum $\mathbf{p}\cdot\de\mathbf{x}=(m_e\mathbf{c}+m_e\bV_e+q_e\bA)\cdot\de\mathbf{x}$. The fact that the electron mean flow (moving with $\bV_e$) does not affect  momentum circulation is explained in  Section \ref{MeanInertia}.

}

\bigskip

%\newpage

\end{document}